\documentclass[aps,twocolumn,showpacs,pra]{revtex4}
\usepackage{amsmath}
\usepackage{amsfonts}
\usepackage{amssymb}
\usepackage{epsfig}
\usepackage{graphicx}
\begin{document}
\title{{Strong Mechanical Squeezing and its Detection}}
\author{G. S. Agarwal$^{1}$ and Sumei Huang$^{2}$}
\affiliation{$^1$ Department of Physics, Oklahoma State University, Stillwater, Oklahoma 74078, USA}
\affiliation{$^2$ Room 105, Building 4, Erqu, Fuzhou Erhua Xincun, Fuzhou, Fujian 350011, China}

\date{\today}
\begin{abstract}
We report an efficient mechanism to generate a squeezed state of a mechanical mirror in an optomechanical system. We use especially tuned parametric amplifier (PA) inside the cavity and the parametric photon phonon processes to transfer quantum squeezing from photons to phonons with almost 100\% efficiency. We get 50\% squeezing of the mechanical mirror which is limited by the PA. We present analytical results for the mechanical squeezing thus enabling one to understand the dependence of squeezing on system parameters like gain of PA, cooperativity, temperature. As in cooling experiments the detrimental effects of mirror's Brownian and zero point noises are strongly suppressed by the pumping power. By judicious choice of the phases, the cavity output is squeezed only if the mirror is squeezed thus providing us a direct measure of the mirror's squeezing. Further considerable larger squeezing of the mirror can be obtained by adding the known feedback techniques.
\end{abstract}
\maketitle

\section{Introduction}

Cavity optomechanics is based on the radiation pressure interaction between light and mechanical resonators at macroscopic scales \cite{Aspelmeyer14}. Currently, with the rapid progress of
practical technologies in cavity optomechanics, the mechanical resonator can be cooled down close to the quantum ground state \cite{Cleland,Teufel,Chan}. Thus it is possible to explore quantum effects in macroscopic systems, including superposition state \cite{Bose, Simon}, entanglement \cite{Kim,Plenio,Agarwal09}, squeezing of light \cite{Fabre,Mancini,Regal,FlorianNPJ,Agarwal14,Qu,Kilda}, squeezing of the mechanical resonator \cite{Agarwal91,Dodonov,Braginsky,Mari,Li,Liao,Kronwald,Zoller,Sumei,Ruskov,Florian,Bowen, Vitali14, Nori, Vanner, Benito, Wollman, PRL, PRX, Bowen1, Pontin, Vinante}, etc.

The quantum squeezing of mechanical modes is important as it can be
used to improve the precision of quantum measurements \cite{Cave}. There
have been many theoretical proposals for generating squeezing of the mechanical mode \cite{Agarwal91,Dodonov,Braginsky,Mari, Li, Liao,Kronwald,Zoller,Sumei,Vitali14,Nori,Vanner,Benito,Ruskov,Florian}. Several experiments have reported squeezing of the mirror to different degrees. Since mechanical motion is represented by an oscillator, the most direct way to produce squeezing is via the well known methods used to squeeze the oscillator motion. One of the early proposals was to modulate the frequency of the oscillator \cite{Agarwal91,Dodonov,Braginsky}. While this is the simplest, it is not easy to adopt for many different kinds of mechanical systems currently in use. Alternate methods to overcome this limitation have been suggested. These include modulation of the external laser \cite{Mari, Li, Liao}; use of a two tone drive one red detuned and the other blue detuned \cite{Kronwald}. One can use a broad band squeezed optical field and couple it into an optomechanical cavity to transfer optical squeezing into mechanical squeezing \cite{Zoller,Sumei}. This method works very well and more than 50\% squeezing of the mirror can be obtained \cite{Zoller,Sumei}. This requires efficient coupling and a highly squeezed broad band field and thus has its own limitations. A more direct way is to have a parametric amplifier placed inside the optomechanical cavity so that the squeezing of the cavity field is generated inside the cavity. These squeezed cavity photons can interact directly with the red-detuned pump laser to produce squeezing of the mechanical mode. This is the main theme of the present work. The degree of the mechanical squeezing will be limited by the squeezing produced by the PA. However one can use the previously used methods like the single quadrature feedback scheme \cite{Vinante} or the weak measurement \cite{Bowen,Pontin} to substantially increase the mirror's squeezing.

While we concentrate on optomechanical couplings linear in mirror’s displacement, the squeezing of the mirror in quadratically coupled OMS has been investigated. In this case one can use a bang-bang technique to kick the mirror mode \cite{Vitali14,Vanner}; use the Duffing nonlinearity \cite{Nori}; use two tone driving \cite{Benito}.

In this paper, we propose a scheme to generate the momentum squeezing of the movable mirror
by placing a degenerate PA inside a Fabry-Perot cavity with one moving mirror. The PA is pumped at twice
the frequency of the anti-Stokes sideband of the driving laser interacting the movable mirror. It is shown that the
squeezing of the cavity field induced by the PA can be transferred to the movable mirror. The achieved momentum squeezing of the mirror depends on the parametric gain, the parametric phase, the power of the input laser,
and the temperature of the environment.

The paper is organized as follows. In Sec. II, we describe
the model, give the quantum Langevin equations, and the steady-state mean values. In Sec. III, we linearize the quantum Langevin equations, derive the stability conditions, calculate the square fluctuations in position and momentum of the movable mirror. In Sec. IV, we discuss how the momentum
squeezing of the movable mirror can be realized by using the PA inside
the cavity. In Sec. V, we derive the analytical expression of the mean square fluctuation in the momentum of the movable mirror. In Sec. VI, we show how the mechanical squeezing can be measured by the output field. Our conclusions are given in Sec. VII.
\section{Model}

\begin{figure}[!h]
\begin{center}
\scalebox{1}{\includegraphics{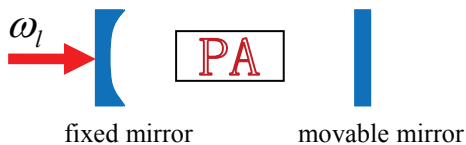}}
\caption{\label{Fig1} Sketch of the optomechanical system to prepare the movable mirror in a squeezed state. A PA is placed inside the cavity, and the pump of the PA is not shown.}
\end{center}
\end{figure}

\begin{figure}[!h]
\begin{center}
\scalebox{0.55}{\includegraphics{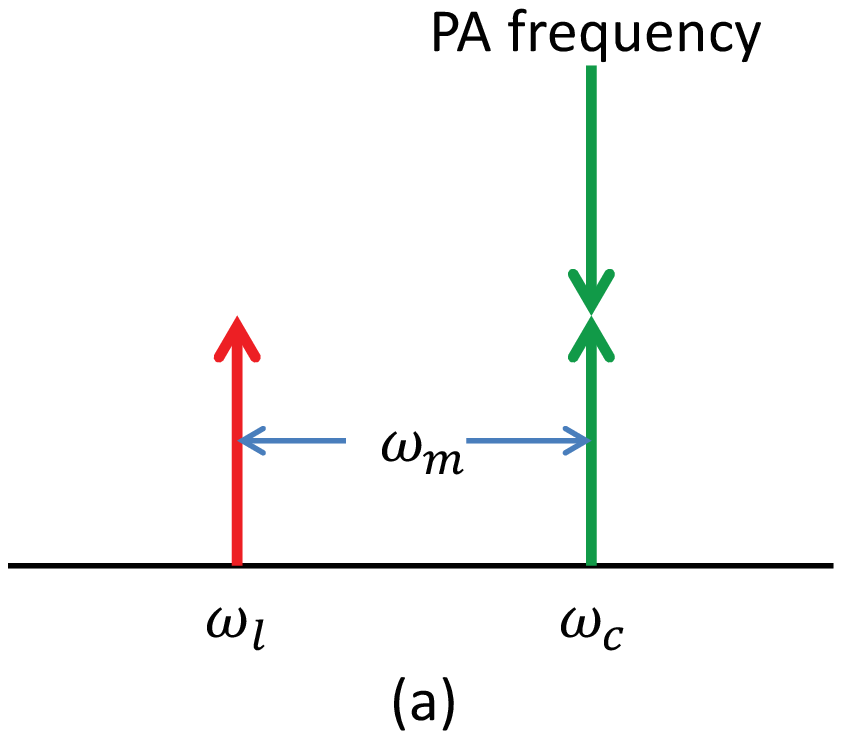}}
\scalebox{0.5}{\includegraphics{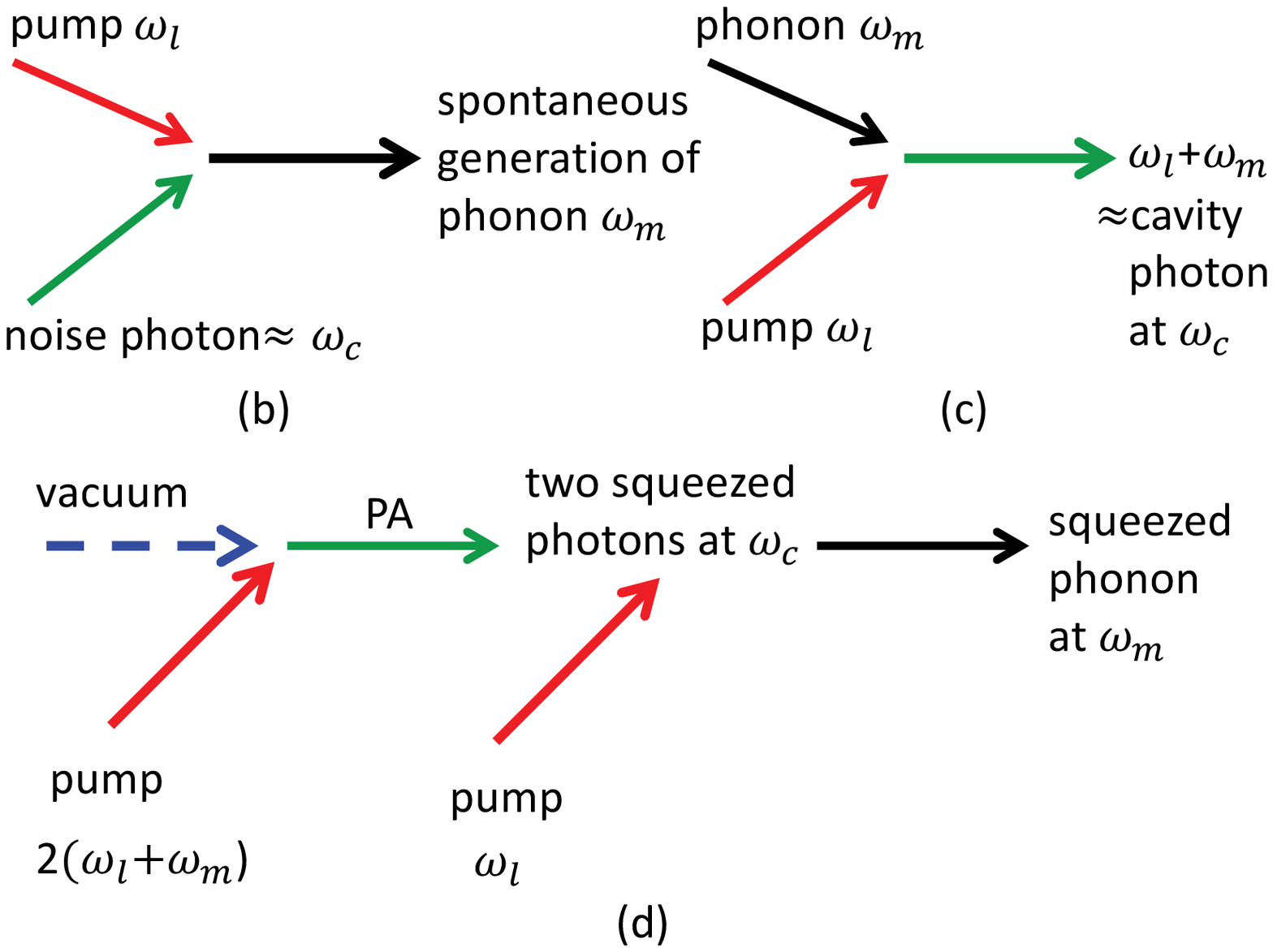}}
\caption{\label{Fig2} Sketch of the physical process. The pump $\omega_{l}$ is red detuned with $\Delta=\omega_{m}$. The system is worked in the resolved sideband limit.}
\end{center}
\end{figure}

We consider a degenerate PA contained in a Fabry-Perot cavity with one fixed mirror and one movable mirror, as shown in Fig. \ref{Fig1}. A degenerate parametric amplifier (PA) is generally used to produce a squeezed light \cite{Walls,Kimble}. We have shown earlier
that a PA inside an optomechanical system can improve the cooling of the movable mirror \cite{Sumei1}. It can also make the observation of the normal-mode splitting \cite{Girvin, Aspelmeyer} of the movable mirror and the output field more accessible \cite{Sumei2}.
The fixed mirror is partially transmitting, while the movable mirror is totally reflecting.
The separation between the two mirrors is $L$. A cavity field with resonance frequency $\omega_{c}$ is driven by an external laser with frequency $\omega_{l}$ and amplitude $\varepsilon_l$. The intracavity photons exert a radiation pressure force on the movable mirror, causing the optomechanical interaction between the cavity field and the movable mirror. Meanwhile, the movable mirror is in contact with a thermal bath in equilibrium at temperature
$T$, which induces a thermal Langevin force acting on the movable mirror. Under the action of these two forces, the mirror makes small oscillations around its equilibrium position. The movable
mirror is treated as a quantum-mechanical harmonic oscillator with effective mass $m$, frequency $\omega_{m}$, and energy decay rate $\gamma_{m}$. In the degenerate PA, we assume that a pump field at frequency $2(\omega_{l}+\omega_{m})$ interacts with a second-order nonlinear optical crystal, thus the signal
and the idler have the same frequency $\omega_{l}+\omega_{m}$. We assume that the gain of the PA is $G$, depending on the power of the pump driving the PA, the phase of the pump driving the PA is $\theta$. The Hamiltonian of the system in the rotating frame at the laser frequency $\omega_{l}$ is given by
\begin{eqnarray}\label{1}
H&=&\hbar(\omega_c-\omega_l)c^{\dag}c-\hbar g_{0} c^{\dag}c(b+b^{\dag})+\hbar\omega_{m}(b^{\dag}b+\frac{1}{2})\nonumber\\
& &+i\hbar\varepsilon_{l} (c^{\dag}-c)+i\hbar G(e^{i\theta}c^{\dag 2}e^{-2i\omega_{m}t}\nonumber\\
& &-e^{-i\theta}c^{2}e^{2i\omega_{m}t}),
\end{eqnarray}
where $c$ and $c^{\dag}$ are the annihilation and creation operators of the cavity mode, satisfying the commutator relation
$[c, c^{\dag}] = 1$, $b$ and $b^{\dag}$ are the annihilation and creation operators of the mechanical mode, satisfying $[b, b^{\dag}] = 1$. The optomechanical interaction strength is $g_{0}=\frac{\omega_{c}}{L}\sqrt{\frac{\hbar}{2m\omega_{m}}}$ in unit of Hz, where $\sqrt{\frac{\hbar}{2m\omega_{m}}}$ is the zero point motion of the movable mirror. The $\varepsilon_{l}$ is related to the power $\wp$ of the laser by $\varepsilon_{l}=\sqrt{\frac{2\kappa\wp}{\hbar\omega_{l}}}$  with $\kappa$ being the cavity decay rate due to the leakage of photons through the partially transmitting mirror. In Eq. (\ref{1}), the first and third terms describe the energies of the optical mode and the mechanical mode, respectively, the second term describes the linear optomechanical coupling between the cavity field and the movable mirror, depending on the photon number $c^{\dag}c$ in the cavity, the fourth term gives the driving of the input laser, the last term represents the coupling between the cavity field and the PA. The physical process can be illustrated in Fig. 2. Fig. \ref{Fig2}(a) shows the frequency relation among the pump photon at frequency $\omega_{l}$, the cavity photon at frequency $\omega_{c}$, the squeezed photon at frequency $\omega_{c}$ from the PA, and the phonon at frequency $\omega_{m}$. Fig. \ref{Fig2}(b) shows that a phonon at frequency $\omega_{m}$ is spontaneously created by a red-detuned pump photon at frequency $\omega_{l}$ interacting with an input noise photon at frequency $\omega_{c}$. Fig. \ref{Fig2}(c) shows that a cavity photon at frequency $\omega_{c}$ is produced when a red-detuned pump photon at frequency $\omega_{l}$ interacting with a phonon at frequency $\omega_{m}$. Fig. \ref{Fig2}(d) shows that a squeezed phonon at frequency $\omega_{m}$ is generated when a red-detuned pump photon at frequency $\omega_{l}$ interacts with a squeezed photon at frequency $\omega_{l}+\omega_{m}\approx\omega_{c}$ from the PA.

According to Heisenberg motion equation and considering the quantum and thermal noises, we obtain the quantum
Langevin equations
\begin{eqnarray}\label{2}
\dot{b}&=&ig_{0}c^{\dag}c-i\omega_m b-\frac{\gamma_m}{2}b+\sqrt{\gamma_m}b_{in},\nonumber\\
\dot{c}&=&-i(\omega_c-\omega_l)c+ig_{0}c(b+b^{\dag})+\varepsilon_{l}+2Ge^{i\theta}c^{\dag}e^{-2i\omega_{m}t}\nonumber\\
& &-\kappa c+\sqrt{2\kappa}c_{in}.
\end{eqnarray}
Here $b_{in}$ is the boson annihilation operator of the thermal noise with zero mean
value, its nonzero correlation functions are
\begin{eqnarray}\label{3}
\langle b_{in}^{\dag}(t)b_{in}(t')\rangle=n^{th}_{m}\delta(t-t'),\nonumber\\
\langle b_{in}(t)b_{in}^{\dag}(t')\rangle=(n^{th}_{m}+1)\delta(t-t'),
\end{eqnarray}
where $n^{th}_{m}=[\exp{[\hbar\omega_{m}/(K_B T)]}-1]^{-1}$ is the initial mean thermal excitation
number in the movable mirror, $K_B$ is the Boltzmann constant.
Moreover, $c_{in}$ is the input quantum noise operator with zero mean
value, its nonzero correlation function is
\begin{eqnarray}\label{4}
\langle c_{in}^{\dag}(t)c_{in}(t')\rangle=n^{th}_{c}\delta(t-t'),\nonumber\\
\langle c_{in}(t)c_{in}^{\dag}(t')\rangle=(n^{th}_{c}+1)\delta(t-t'),
\end{eqnarray}
where $n^{th}_{c}=[\exp{[\hbar\omega_{c}/(K_B T)]}-1]^{-1}$ is the initial mean thermal excitation
number in the optical mode.
The steady state mean values of the system operators are
\begin{eqnarray}\label{5}
c_{s}&=&\frac{\varepsilon_{l}}{\kappa+i\Delta},\nonumber\\
b_{s}&=&\frac{ig_{0}|c_{s}|^2}{\frac{\gamma_{m}}{2}+i\omega_{m}},
\end{eqnarray}
where $\Delta=\omega_{c}-\omega_{l}-g_{0}(b_{s}+b_{s}^{*})$ is the effective cavity detuning from the frequency of the input laser in the presence of the radiation pressure, depending on the mechanical motion. The $c_{s}$ is the steady-state amplitude of the cavity field, $b_{s}$ determines the steady-state displacement of the movable mirror. The mean numbers of the cavity photons and the mechanical phonons are given by $|c_{s}|^2$ and $|b_{s}|^2$, respectively.

\section{Radiation pressure and quantum fluctuations}
In order to show the movable mirror in a squeezed state, we need to calculate the position and momentum fluctuations of the movable mirror.
Here we are interested in the strong-driving regime so that the intracavity photon number $|c_{s}|^2$ satisfies $|c_{s}|^2\gg1$. Let $b=b_{s}+\delta b$ and $c=c_{s}+\delta c$, where $\delta b$ and $\delta c$ are the small fluctuation operators around the steady state mean values, thus Eq. (2) can be linearized by neglecting higher than first order terms in the fluctuations \cite{Pina}. Introducing the slow varying fluctuation operators by $\delta b=\delta \tilde{b}e^{-i\omega_{m}t}$, $\delta c=\delta \tilde{c}e^{-i\Delta t}$, $b_{in}=\tilde{b}_{in}e^{-i\omega_{m}t}$, $c_{in}=\tilde{c}_{in}e^{-i\Delta t}$, we obtain the linearized quantum Langevin equations
\begin{eqnarray}\label{6}
\delta\dot{\tilde{b}}&=&i[g^{*}\delta \tilde{c}e^{-i(\Delta-\omega_{m})t}+g\delta \tilde{c}^{\dag}e^{i(\Delta+\omega_{m})t}]-\frac{\gamma_m}{2}\delta \tilde{b}\nonumber\\
& &+\sqrt{\gamma_m}\tilde{b}_{in},\nonumber\\
\delta\dot{\tilde{c}}&=&-\kappa\delta \tilde{c}+ig[\delta \tilde{b}e^{-i(\omega_{m}-\Delta)t}+\delta \tilde{b}^{\dag}e^{i(\omega_{m}+\Delta)t}]\nonumber\\
& &+2Ge^{i\theta}\delta  \tilde{c}^{\dag}e^{2i(\Delta-\omega_{m})t}+\sqrt{2\kappa}\tilde{c}_{in},
\end{eqnarray}
where $g=g_{0}c_{s}$ is the effective optomechanical coupling rate, depending on the power $\wp$ of the input laser.
We assume that the driving field is red-detuned from the cavity resonance ($\Delta=\omega_{m}$), thus the anti-Stokes
scattered light is nearly resonant with the cavity field. And we assume that the system is working in the resolved sideband limit $\omega_{m}\gg\kappa$, the mechanical quality factor is high $\omega_{m}\gg\gamma_{m}$, the mechanical frequency $\omega_{m}$ is much larger than $|g|$ and $2G$. Under these conditions, the rotating wave approximation can be made, the fast
oscillating term $e^{2i\omega_{m}t}$ in Eq. (\ref{6}) can be ignored, Eq. (\ref{6}) can be simplified to
\begin{eqnarray}\label{7}
\delta\dot{\tilde{b}}&=&ig^{*}\delta \tilde{c}-\frac{\gamma_m}{2}\delta \tilde{b}+\sqrt{\gamma_m}\tilde{b}_{in},\nonumber\\
\delta\dot{\tilde{c}}&=&-\kappa\delta \tilde{c}+ig\delta \tilde{b}+2Ge^{i\theta}\delta \tilde{c}^{\dag}+\sqrt{2\kappa}\tilde{c}_{in}.
\end{eqnarray}
Introducing the position and momentum fluctuations of the mechanical oscillator as $\delta Q=\frac{1}{\sqrt{2}}(\delta\tilde{b}+\delta\tilde{b}^{\dag})$ and $\delta P=\frac{1}{\sqrt{2}i}(\delta\tilde{b}-\delta\tilde{b}^{\dag})$, and the amplitude and phase fluctuations of the cavity field as $\delta x=\frac{1}{\sqrt{2}}(\delta
\tilde{c}+\delta\tilde{c}^{\dag})$ and $\delta y =\frac{1}{\sqrt{2}i}(\delta \tilde{c}-\delta \tilde{c}^{\dag})$, the amplitude and phase fluctuations of the input quantum noise as $ x_{in}=\frac{1}{\sqrt{2}}(
\tilde{c}_{in}+\tilde{c}_{in}^{\dag})$ and $y_{in}=\frac{1}{\sqrt{2}i}(\tilde{c}_{in}-\tilde{c}_{in}^{\dag})$, and the position and momentum fluctuations of the thermal noise as $Q_{in}=\frac{1}{\sqrt{2}}(\tilde{b}_{in}+\tilde{b}_{in}^{\dag})$ and $P_{in}=\frac{1}{\sqrt{2}i}(\tilde{b}_{in}-\tilde{b}_{in}^{\dag})$,
the equation (\ref{7}) can be written as the matrix form
\begin{equation}\label{8}
\dot{f}(t)=Mf(t)+n(t),
\end{equation}
where $f(t)$ is the column vector of the fluctuations, and $n(t)$
is the column vector of the noise sources. Their transposes are
\begin{eqnarray}\label{9}
f(t)^{T}&=&(\delta Q,\delta P,\delta x,\delta y),\nonumber\\
n(t)^{T}&=&(\sqrt{\gamma_{m}}Q_{in},\sqrt{\gamma_{m}}P_{in},\sqrt{2\kappa} x_{in},\sqrt{2\kappa} y_{in});
\end{eqnarray}
and the matrix $M$ is given by
\begin{widetext}
\begin{equation}\label{10}
M=\left(
  \begin{array}{cccc}
    -\frac{\gamma_{m}}{2} & 0 & \frac{i}{2}(g^{*}-g) & -\frac{1}{2}(g+g^{*}) \\
    0 & -\frac{\gamma_{m}}{2} & \frac{1}{2}(g+g^{*}) & \frac{i}{2}(g^{*}-g) \\
    \frac{i}{2}(g-g^{*})  & -\frac{1}{2}(g+g^{*})  & -(\kappa-2G\cos\theta) & 2G\sin\theta \\
     \frac{1}{2}(g+g^{*})  & \frac{i}{2}(g-g^{*}) & 2G\sin\theta & -(\kappa+2G\cos\theta) \\
  \end{array}
\right).
\end{equation}
\end{widetext}
The stability conditions of the system can be obtained by requiring that all the eigenvalues of
the matrix $M$ have negative real parts. Applying the Routh-Hurwitz
criterion \cite{stable1,stable2}, we find the stability conditions
\begin{eqnarray}\label{11}
& &\frac{1}{4} \gamma_{m}^3+2\kappa(\kappa^2-4G^2)+(2\kappa+\gamma_{m})(|g|^2+2\kappa\gamma_{m})>0,\nonumber\\
& &2\kappa\gamma_{m}(\kappa^2-4G^2)^2+\big[(2\kappa+\gamma_{m})^2|g|^2+(4\kappa+\gamma_{m})\kappa\gamma_{m}^2\big]\nonumber\\
& &\quad\times(\kappa^2-4G^2)+\frac{\gamma_{m}^3}{4}\big[\frac{\kappa\gamma_{m}^2}{2}+(2\kappa+\gamma_{m})|g|^2\big]\nonumber\\
& &\quad+\kappa\gamma_{m}(2\kappa+\gamma_{m})\big[\kappa\gamma_{m}^2+(2\kappa+\frac{3}{2}\gamma_{m})|g|^2\big]>0,\nonumber\\
& &\frac{1}{4} \gamma_{m}^2(\kappa^2-4G^2)+|g|^2(|g|^2+\kappa\gamma_{m})>0.
\end{eqnarray}
 Note that the stability conditions are independent
of the parametric phase $\theta$. The system stays in the stable regime only if $G<0.5\kappa$.

By taking the Fourier transform of Eq. (\ref{8}) and solving it in the frequency domain, we obtain the expressions for the position and momentum
fluctuations of the movable mirror
\begin{eqnarray}\label{12}
\delta Q(\omega)&=&A_{1}(\omega)x_{in}(\omega)+B_{1}(\omega)y_{in}(\omega)+E_{1}(\omega)Q_{in}(\omega)\nonumber\\
& &+F_{1}(\omega)P_{in}(\omega),\nonumber\\
\delta P(\omega)&=&A_{2}(\omega)x_{in}(\omega)+B_{2}(\omega)y_{in}(\omega)+E_{2}(\omega)Q_{in}(\omega)\nonumber\\
& &+F_{2}(\omega)P_{in}(\omega),
\end{eqnarray}
where
\begin{eqnarray}\label{13}
A_{1}(\omega)&=&\frac{\sqrt{2\kappa}i}{d(\omega)}\Big\{v(\omega)\big[G\alpha-iu(\omega)\text{Im}(g)\big]-i|g|^2\text{Im}(g)\Big\},\nonumber\\
B_{1}(\omega)&=&\frac{\sqrt{2\kappa}}{d(\omega)}\Big\{v(\omega)\big[G\beta-u(\omega)\text{Re}(g)\big]-|g|^2\text{Re}(g)\Big\},\nonumber\\
E_{1}(\omega)&=&\frac{\sqrt{\gamma_{m}}}{d(\omega)}\Big\{\big[u(\omega)^2-4G^2\big]v(\omega)+|g|^2u(\omega)+G\Gamma\Big\},\nonumber\\
F_{1}(\omega)&=&\frac{\sqrt{\gamma_{m}}}{d(\omega)} iG(g^2e^{-i\theta}-g^{*2}e^{i\theta}),\nonumber\\
A_{2}(\omega)&=&\frac{\sqrt{2\kappa}}{d(\omega)}\Big\{v(\omega)\big[G\beta+u(\omega)\text{Re}(g)\big]+|g|^2\text{Re}(g)\Big\},\nonumber\\
B_{2}(\omega)&=&-\frac{\sqrt{2\kappa}i}{d(\omega)}\Big\{v(\omega)\big[G\alpha+iu(\omega)\text{Im}(g)\big]+i|g|^2\text{Im}(g)\Big\},\nonumber\\
E_{2}(\omega)&=&\frac{\sqrt{\gamma_{m}}}{d(\omega)} iG(g^2e^{-i\theta}-g^{*2}e^{i\theta}),\nonumber\\
F_{2}(\omega)&=&\frac{\sqrt{\gamma_{m}}}{d(\omega)}\Big\{\big[u(\omega)^2-4G^2\big]v(\omega)+|g|^2u(\omega)-G\Gamma\Big\},\nonumber\\
\end{eqnarray}
with $\alpha=e^{i\theta}g^{*}-e^{-i\theta}g$, $\beta=e^{i\theta}g^{*}+e^{-i\theta}g$, $\Gamma=g^2e^{-i\theta}+g^{*2}e^{i\theta}$, $v(\omega)=\frac{\gamma_{m}}{2}-i\omega$, $u(\omega)=\kappa-i\omega$, and
\begin{equation}\label{14}
d(\omega)=\big[u(\omega)v(\omega)+|g|^2\big]^2-4G^2v(\omega)^2.
\end{equation}
In Eq. (\ref{12}), the first two terms in $\delta Q(\omega)$ and $\delta P(\omega)$ are from the radiation pressure contribution, the last two terms are from the thermal noise contribution. In the absence of the optomechanical coupling $(g=0)$, the movable mirror makes quantum Brownian motion because of the coupling to the environment, $\delta Q(\omega)=\frac{\sqrt{\gamma_{m}}}{\frac{\gamma_{m}}{2}-i\omega}Q_{in}$, $\delta P(\omega)=\frac{\sqrt{\gamma_{m}}}{\frac{\gamma_{m}}{2}-i\omega}P_{in}$.
The spectra of fluctuations in the position and momentum of the movable mirror are defined by
\begin{eqnarray}\label{15}
S_{Z}(\omega)&=&\frac{1}{4\pi}\int_{-\infty}^{+\infty} d\Omega\ e^{-i(\omega+\Omega)t}[\langle \delta Z(\omega)\delta Z(\Omega)\rangle\nonumber\\
& &+\langle\delta Z(\Omega)\delta Z(\omega)\rangle],\qquad Z=Q, P.
\end{eqnarray}
By the aid of the nonzero correlation functions of the noise sources in the frequency domain,
\begin{eqnarray}\label{16}
\langle Q_{in}(\omega)Q_{in}(\Omega)\rangle&=&\langle P_{in}(\omega)P_{in}(\Omega)\rangle\nonumber\\
&=&(n^{th}_{m}+\frac{1}{2})2\pi\delta(\omega+\Omega),\nonumber\\
\langle Q_{in}(\omega)P_{in}(\Omega)\rangle&=&-\langle P_{in}(\omega)Q_{in}(\Omega)\rangle=\frac{i}{2}2\pi\delta(\omega+\Omega),\nonumber\\
\langle x_{in}(\omega)x_{in}(\Omega)\rangle&=&\langle y_{in}(\omega)y_{in}(\Omega)\rangle\nonumber\\
&=&(n^{th}_{c}+\frac{1}{2})2\pi\delta(\omega+\Omega),\nonumber\\
\langle x_{in}(\omega)y_{in}(\Omega)\rangle&=&-\langle y_{in}(\omega)x_{in}(\Omega)\rangle=\frac{i}{2}2\pi\delta(\omega+\Omega),\nonumber\\
\end{eqnarray}
we obtain the spectra of fluctuations in the position and momentum of the movable
mirror
\begin{eqnarray}\label{17}
S_{Q}(\omega)&=&[A_{1}(\omega)A_{1}(-\omega)+B_{1}(\omega)B_{1}(-\omega)](n^{th}_{c}+\frac{1}{2})\nonumber\\
& &+[E_{1}(\omega)E_{1}(-\omega)+F_{1}(\omega)F_{1}(-\omega)](n^{th}_{m}+\frac{1}{2}),\nonumber\\
S_{P}(\omega)&=&[A_{2}(\omega)A_{2}(-\omega)+B_{2}(\omega)B_{2}(-\omega)](n^{th}_{c}+\frac{1}{2})\nonumber\\
& &+[E_{2}(\omega)E_{2}(-\omega)+F_{2}(\omega)F_{2}(-\omega)](n^{th}_{m}+\frac{1}{2}),\nonumber\\
\end{eqnarray}
where the first term proportional
to $n^{th}_{c}+\frac{1}{2}$ in $S_{Q}(\omega)$ and $S_{P}(\omega)$ is from the radiation pressure contribution, while the second term proportional
to $n^{th}_{m}+\frac{1}{2}$ is from
the thermal noise contribution. In the absence of the cavity field, the spectra of fluctuations in position and momentum of the movable mirror are given by
$S_{Q}(\omega)=S_{P}(\omega)=\frac{\gamma_{m}}{\frac{\gamma_{m}^2}{4}+\omega^2}(n^{th}_{m}+\frac{1}{2})$, whose peaks are located at frequency zero with full width $\gamma_{m}$ at half maximum.
The mean square fluctuations $\langle \delta Q(t)^2\rangle$ and $\langle \delta P(t)^2\rangle$ in the position and momentum of the movable mirror are
determined by
\begin{eqnarray}\label{18}
\langle \delta Z(t)^2\rangle&=&\frac{1}{2\pi}\int^{+\infty}_{-\infty} d\omega\ S_{Z}(\omega),\quad Z=Q, P.
\end{eqnarray}
Without the optomechanical coupling, we find $\langle \delta Q(t)^2\rangle=\langle \delta P(t)^2\rangle=n^{th}_{m}+\frac{1}{2}$. For $T=0$ K, the movable mirror is in the ground state $(n^{th}_{m}=0)$, $\langle \delta Q(t)^2\rangle=\langle \delta P(t)^2\rangle=\frac{1}{2}$.
According to the Heisenberg uncertainty principle, the product of the mean square fluctuations $\langle \delta Q(t)^2\rangle$ and $\langle \delta P(t)^2\rangle$ satisfies the following inequality
\begin{eqnarray}\label{19}
 \langle \delta Q(t)^2\rangle\langle \delta P(t)^2\rangle\geq|\frac{1}{2}[Q,P]|^2,
\end{eqnarray}
where $[Q,P]=i$.
If either $\langle \delta Q(t)^2\rangle$ or $ \langle \delta P(t)^2\rangle$ is below $\frac{1}{2}$, the state of the movable mirror exhibits quadrature squeezing. The degree of the squeezing can be expressed in the dB unit, which can be calculated by
$-10\log_{10}\frac{\langle \delta P(t)^2\rangle}{\langle \delta P(t)^2\rangle_{vac}}$
with $\langle \delta P(t)^2\rangle_{vac}$ being the momentum variance of the vacuum state and $\langle \delta P(t)^2\rangle_{vac}=\frac{1}{2}$.

\section{The mechanical squeezing}
In this section, we numerically evaluate the mean
square fluctuations in the position and momentum of the
movable mirror given by Eq. (\ref{18}) to show quadrature squeezing of
the movable mirror under the action of the PA. The values of the parameters are chosen to be similar to those in the experiment demonstrating mechanical squeezing with two pumps \cite{Wollman}: $\omega_{m}/\kappa=10$, $\gamma_{m}/\kappa=10^{-5}$, the mechanical quality factor $Q'=\omega_{m}/\gamma_{m}=10^{6}$. For convenience, we define the optomechanical cooperativity parameter $C=|g|^2/(\kappa\gamma_{m})$, which is proportional to the power $\wp$ of the external laser. From the numerical results, it is found that $\langle \delta Q(t)^2\rangle$ can not be less than $\frac{1}{2}$,
but $\langle \delta P(t)^2\rangle$ can be less than $\frac{1}{2}$. Therefore we
focus on discussing $\langle \delta P(t)^2\rangle$ here.

\begin{figure}[htp]
\begin{center}
\scalebox{0.95}{\includegraphics{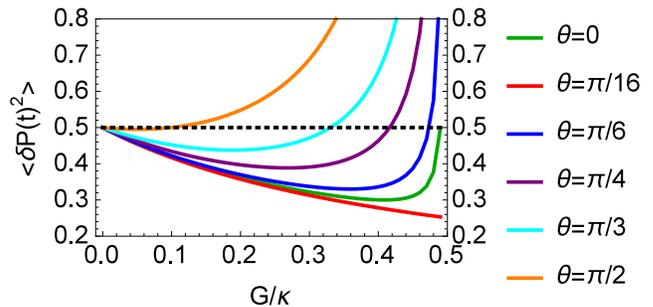}}
\caption{\label{Fig3}  The mean square fluctuation $\langle \delta P(t)^2\rangle$
versus the parametric gain $G$ for different parametric phases $\theta=0, \pi/16, \pi/6, \pi/4, \pi/3, \pi/2$ when $C=400$ and $T=0$ K. The flat dotted line
represents the momentum variance of the vacuum state $\langle \delta P(t)^2\rangle_{vac}=0.5$. }
\end{center}
\end{figure}

The mean square fluctuation $\langle \delta P(t)^2\rangle$ as a function of the parametric gain $G$ for different parametric phases $\theta=0, \pi/16, \pi/6, \pi/4, \pi/3, \pi/2$ when $C=400$ and $T=0$ K is shown in Fig. \ref{Fig3}. When $C=400$, $\kappa=5\sqrt{10}|g|$, the system is in the weak-coupling regime, and the conditions for the rotating wave approximation are satisfied. From Fig. \ref{Fig3}, it is seen that $\langle \delta P(t)^2\rangle$=0.5 in the absence of the PA ($G=0$), thus there is no squeezing in the momentum fluctuation of the movable mirror. In the presence of the PA ($G\neq0$), $\langle \delta P(t)^2\rangle$ can be less than 0.5 except $\theta=\frac{\pi}{2}$. Hence the addition of the PA in the optomechanical system can realize the momentum squeezing of the movable mirror. Furthermore, it is observed that the minimum value of $\langle \delta P(t)^2\rangle$ is the smallest when $\theta=\pi/16$, which is $\langle \delta P(t)^2\rangle\approx0.253$ at $G=0.49\kappa$, the corresponding amount of the maximum momentum squeezing is about 49.4\%, the degree of the squeezing is about $2.96$ dB. The squeezing of the cavity field in the absence of the optomechanical coupling is given in the Appendix. The maximum phase squeezing of the cavity field is about $2.96$ dB when $G=0.49\kappa$ and $\theta=0$. Note that the maximum momentum squeezing of the movable mirror is equal to the maximum phase squeezing of the cavity field, but they happen at different parametric phases $\theta$. The phase difference $\pi/16$ is related to the phase of $g^2$ and $\arctan[-g^{2}]=\pi/16$. Thus the squeezing of the cavity field is totally transferred into the movable mirror. This is because driving the system by the red-detuned laser $\Delta=\omega_{m}$ in the resoved sideband limit makes the optomechanical interaction between the movable mirror and the cavity field like a beamsplitter interaction.

\begin{figure}[htp]
\begin{center}
\scalebox{0.8}{\includegraphics{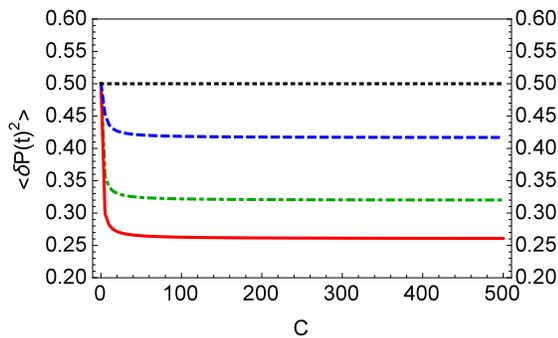}}
\caption{\label{Fig4}  The mean square fluctuation $\langle \delta P(t)^2\rangle$
versus the cooperativity parameter $C$ for different parametric phases $\theta=0$ (dotdashed), $\pi/16$ (solid), $\pi/6$ (dashed) when $G=0.46\kappa$ and $T=0$ K. The flat dotted line
represents the momentum variance of the vacuum state $\langle \delta P(t)^2\rangle_{vac}=0.5$. }
\end{center}
\end{figure}

The mean square fluctuation $\langle \delta P(t)^2\rangle$ as a function of the cooperativity parameter $C$ for different parametric phases $\theta=0, \pi/16, \pi/6$ when $G=0.46\kappa$ and $T=0$ K is shown in Fig. \ref{Fig4}. In the absence of the optomechanical coupling ($C=0$) between the cavity field and the movable mirror, it is seen that $\langle \delta P(t)^2\rangle=0.5$. In the presence of the optomechanical coupling ($C\neq0$), $\langle \delta P(t)^2\rangle$ drops to about 0.320, 0.261, 0.417 for $\theta=0, \pi/16, \pi/6$, respectively, thus the optomechanical coupling can lead to the momentum squeezing of the movable mirror. The corresponding degrees of the squeezing are about 1.94 dB, 2.82 dB, 0.79 dB for $\theta=0, \pi/16, \pi/6$, respectively. It is noted that the momentum squeezing of the movable mirror almost keeps constant when the cooperativity parameter $C$ is larger than a certain value and it persists over a very wide range.

In this paragraph, we discuss previous results on mechanical squeezing. The mechanical squeezing is not larger than 3 dB in \cite{Mari, Li, Liao} as in this work. In the current work it is limited by the squeezing that a parametric device can produce. A relatively large mechanical squeezing can be
achieved by feeding in squeezed light \cite{Zoller,Sumei}. Here one gets about 6 dB squeezing by feeding in light with about 9 dB squeezing. The two tone driving as discussed in detail in Ref. \cite{Kronwald} can also produce large squeezing (more than 3 dB). For this, the intensity of the blue-detuned drive has to be close but smaller than the intensity of the red-detuned drive and the cooperativity parameter $C$ has to be large. The latter requirement should not be in conflict with the dropping of the nonresonant terms in the case of two tone driving. In addition, the mechanical squeezing beyond 3 dB can be created by quantum measurement and feedback to remove the effect of the quantum back action \cite{Ruskov,Florian}. Several experiments have reported good mechanical squeezing. The best experimental mechanical squeezing is roughly 1.0 dB in
\cite{Wollman,PRL,PRX}, 6.2 dB in \cite{Bowen1}, 7.4 dB in \cite{Pontin}, and 11.5 dB in \cite{Vinante},
respectively. It is clear that additional methods are to be used to go beyond 3 dB squeezing. This is briefly discussed at the end of Sec. V.

\begin{figure}[htp]
\begin{center}
\scalebox{0.95}{\includegraphics{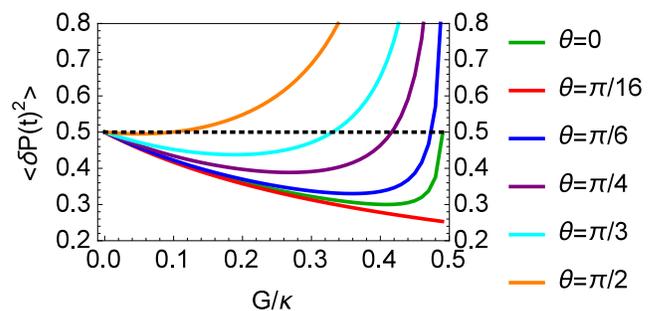}}
\caption{\label{Fig5}  The mean square fluctuation $\langle \delta P(t)^2\rangle$
versus the parametric gain $G$ for different parametric phases $\theta=0, \pi/16, \pi/6, \pi/4, \pi/3, \pi/2$ when $C=400$ and $T=0$ K. The flat dotted line
represents the momentum variance of the vacuum state $\langle \delta P(t)^2\rangle_{vac}=0.5$. }
\end{center}
\end{figure}

We find that the amount of squeezing of the mechanical mirror is not very sensitive to the parameters. We next choose parameters corresponding to an optical cavity. We take $\omega_{m}/\kappa=10$, $\gamma_{m}/\kappa=10^{-3}$.
The mean square fluctuation $\langle \delta P(t)^2\rangle$ as a function of the parametric gain $G$ for different parametric phases $\theta=0, \pi/16, \pi/6, \pi/4, \pi/3, \pi/2$ when $C=400$ and $T=0$ K is shown in Fig. \ref{Fig5}. When $C=400$, $\kappa=\sqrt{10}|g|/2$, the system is in the weak-coupling regime, and the conditions for the rotating wave approximation are satisfied. It is seen that $\langle \delta P(t)^2\rangle$ takes the smallest value 0.253 when $\theta=\pi/16$ and $G=0.49\kappa$, which is similar to that in Fig. \ref{Fig3}.
The mean square fluctuation $\langle \delta P(t)^2\rangle$ as a function of the cooperativity parameter $C$ for different parametric phases $\theta=0, \pi/16, \pi/6$ when $G=0.46\kappa$ and $T=0$ K is similar to Fig. \ref{Fig4}. In the presence of the optomechanical coupling, $\langle \delta P(t)^2\rangle$ drops to about 0.320, 0.261, 0.416 for $\theta=0, \pi/16, \pi/6$, respectively.

\begin{figure}[htp]
\begin{center}
\scalebox{0.8}{\includegraphics{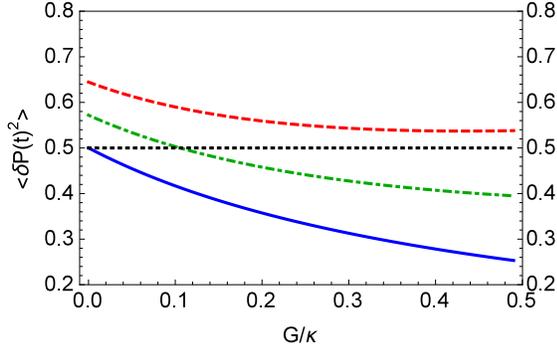}}
\caption{\label{Fig6}  The mean square fluctuation $\langle \delta P(t)^2\rangle$
versus the parametric gain $G$ for different temperatures of the environment $T=$0 K(solid), 10 mK (dotdashed), 20 mK (dashed) when $C=400$ and $\theta=\pi/16$. The flat dotted line
represents the momentum variance of the vacuum state $\langle \delta P(t)^2\rangle_{vac}=0.5$.}
\end{center}
\end{figure}

We next examine the effect of the Brownian noise on squeezing i.e. the effect of the temperature of the environment. We need the values of the cavity frequency $\omega_{c}$ and the mechanical frequency $\omega_{m}$. We assume $\omega_{c}=2\pi\times6.23$ GHz and $\omega_{m}=2\pi\times 3.6$ MHz \cite{Wollman}. The mean square fluctuation $\langle \delta P(t)^2\rangle$ as a function of the parametric gain $G$ for different temperatures of the environment when $C=400$ and $\theta=\pi/16$ is plotted in Fig. \ref{Fig6}. For $T=$0 K, 10 mK, 20 mK, the corresponding initial mean thermal excitation numbers $n^{th}_{c}$ in the optical mode are 0, $1.03\times10^{-13}$, and $3.22\times10^{-7}$, respectively, the corresponding initial mean thermal excitation numbers $n^{th}_{m}$ in the mechanical mode are 0, 57.4, and 115.3, respectively. It is noted that increasing the temperature of the environment would decrease the momentum squeezing of the movable mirror. For example, when $G=0.49\kappa$, $T=0$ K, 10 mK, $\langle\delta P(t)^2\rangle\approx0.253$, 0.395, respectively, the corresponding degrees of the squeezing are about 2.96 dB, 1.02 dB, respectively. When the temperature of the environment is increased to $T=20$ mK, $\langle \delta P(t)^2\rangle$ is always larger than 0.5, thus the squeezing of the mechanical mode does not occur. We have confirmed that for the optical cavity case the results of Fig. 6 hold with almost no change. For brevity we do not present the figure for the optical cavity case.

The PA inside the OM cavity can produce a number of novel effects besides squeezing of the mirror and cooling. Some of these are generation of the  genuine tripartite entangled states \cite{Xuereb}, enhancement of the precision of optomechanical position detection \cite{Peano}, enhancement of the effective optomechanical interaction strength \cite{Lu,Nation}. The latter could become important for getting closer to single photon coupling regime.

\section{ANALYTICAL APPROACH TO UNDERSTAND MECHANICAL SQUEEZING}
In this section, we will present an analytical approach to understand the result of Sec. IV.
In the weakly optomechanical coupling regime $\kappa\gg |g|$, in which the photons leak out of the cavity much faster than
the optomechanical interaction, the cavity field follows the mechanical motion adiabatically. The adiabatical approximation can be made, thus $\delta\dot{\tilde{c}}=0$. We obtain
\begin{eqnarray}\label{20}
\delta \tilde{c}&=&\frac{1}{\kappa^2-4G^2}\big(i\kappa g \delta \tilde{b}-i2Ge^{i\theta}g^{*}\delta \tilde{b}^{\dag}+2Ge^{i\theta}\sqrt{2\kappa}\tilde{c}_{in}^{\dag}\nonumber\\
& &+\kappa\sqrt{2\kappa}\tilde{c}_{in}\big).
\end{eqnarray}
Substituting $\delta \tilde{c}$ into Eq. (\ref{7}), we have
\begin{eqnarray}\label{21}
\delta \dot{\tilde{b}}&=&-\Big(\frac{\kappa|g|^2}{\kappa^2-4G^2}+\frac{\gamma_{m}}{2}\Big)\delta \tilde{b}+\frac{2Ge^{i\theta}g^{*2}}{\kappa^2-4G^2}\delta \tilde{b}^{\dag}\nonumber\\
& &+\frac{ig^{*}\sqrt{2\kappa}}{\kappa^2-4G^2}\Big(2Ge^{i\theta}\tilde{c}_{in}^{\dag}+\kappa \tilde{c}_{in}\Big)+\sqrt{\gamma_{m}}\tilde{b}_{in}.
\end{eqnarray}
In the absence of the PA $(G=0)$ or the cavity field $(g=0)$, it is noted that $\delta \dot{\tilde{b}}$ does not depend on $\tilde{b}^{\dag}$, thus the squeezing of the movable mirror does not appear. In the presence of the PA and the cavity field, $\delta \dot{\tilde{b}}$ depends on $\tilde{b}^{\dag}$. This parametric coupling can lead to the squeezing of the movable mirror. Therefore, the PA in the cavity can realize the squeezing of the movable mirror.

In the parameter domain we are working the term $\gamma_{m}$ in the coefficient of $\delta \tilde{b}$ can be ignored. Let $G_{0}=2G/\kappa$ and we choose a value of $\theta$ such that $g^{*2}e^{i\theta}=-|g|^2$ $(\theta=\pi/16$), then we write (21) as
\begin{eqnarray}\label{22}
\delta \dot{\tilde{b}}&=&-\frac{|g|^2}{(1-G_{0}^{2})\kappa}\delta \tilde{b}-\frac{G_{0}|g|^2}{(1-G_{0}^2)\kappa}\delta \tilde{b}^{\dag}\nonumber\\
& &+\frac{ig^{*}\sqrt{2\kappa}}{(1-G_{0}^2)\kappa}\Big(G_{0}e^{i\theta}\tilde{c}_{in}^{\dag}+\tilde{c}_{in}\Big)+\sqrt{\gamma_{m}}\tilde{b}_{in}.
\end{eqnarray}
From Eq. (\ref{22}), we get the equation for the momentum fluctuation $\delta P$ as
\begin{eqnarray}\label{23}
\delta \dot{P}=-\frac{|g|^2}{\kappa(1+G_{0})}\delta P+h(t)+f(t),
\end{eqnarray}
where the quantum Langevin forces are given by
\begin{eqnarray}
h(t)&=&\frac{\sqrt{\gamma_{m}}}{\sqrt{2}i}(\tilde{b}_{in}-\tilde{b}^{\dag}_{in}),\label{24}\\
f(t)&=&\frac{g^{*}\sqrt{\kappa}}{\kappa(1+G_{0})}(\tilde{c}_{in}-\tilde{c}_{in}^{\dag}e^{i\theta})\label{25}.
\end{eqnarray}
Using Eq. (\ref{25}) and Eq. (\ref{4}), we obtain the correlation function of $f(t)$
\begin{eqnarray}\label{26}
\langle f(t)f(t')\rangle=\frac{|g|^2}{\kappa(1+G_{0})^2}(1+2n_{c}^{th})\delta(t-t').
\end{eqnarray}
The correlation function of $h(t)$ can be calculated using (\ref{24}) and (\ref{3}).
\begin{eqnarray}\label{27}
\langle h(t)h(t')\rangle=\frac{\gamma_{m}}{2}(1+2n_{m}^{th})\delta(t-t').
\end{eqnarray}
We now obtain the equation for $\langle \delta P^2\rangle$ using (\ref{23}), (\ref{26}), and (\ref{27}) as
\begin{eqnarray}\label{28}
\frac{\partial\langle \delta P^2\rangle}{\partial t}&=&-\frac{2|g|^2}{\kappa(1+G_{0})}\langle \delta P^2\rangle+\frac{|g|^2}{\kappa(1+G_{0})^2}(1+2n_{c}^{th})\nonumber\\
& &+\frac{\gamma_{m}}{2}(1+2n_{m}^{th}),
\end{eqnarray}
and therefore we get the analytical result for the squeezing of the quadrature $P$ in the steady state as
\begin{equation}\label{29}
\langle\delta P^2\rangle=\frac{1}{2(1+G_{0})}(1+2n_{c}^{th})+\frac{\gamma_{m}\kappa(1+G_{0})}{4|g|^2}(1+2n_{m}^{th}).
\end{equation}
For $G_{0}\leq1$, $|g|^{2}/(\gamma_{m}\kappa)=400$, we find
\begin{equation}\label{30}
\langle \delta P^2\rangle\approx\frac{1}{4}(1+2n_{c}^{th})+\frac{1}{800}(1+2n_{m}^{th}),
\end{equation}
which gives values about 0.25, 0.40, and 0.55 for $T=0$ ($n_{c}^{th}=n_{m}^{th}=0$), 10 mK ($n_{c}^{th}=1.03\times10^{-13}$, $n_{m}^{th}=57.4$), and 20 mK ($n_{c}^{th}=3.22\times10^{-7}$, $n_{m}^{th}=115.3$), respectively. These analytical results are in excellent agreement with the numerical results in Fig. 6 for $G/\kappa$ close to but less than 0.5. A very important feature of the result (\ref{29}) which is to noticed is the suppression of the Brownian noise by the cooperativity parameter $C$. As we have mentioned earlier and as has been realized by several others \cite{Ruskov,Florian}, the 3 dB limit can be broken by using the feedback mechanism as in Ref. \cite{Vinante}. Let $\eta$ be the dimensionless feedback gain parameter, then detailed calculations show that the squeezing given by Eq. (\ref{29}) is reduced by a factor of $[1+\frac{1}{2C}(1+G_{0})(1+\frac{\eta}{2})]$. The maximum value of $\eta$ is limited by the stability of the dynamical equations. Thus an order to get 75\% squeezing (6 dB), we need the condition $\eta/C\sim2$. Still larger squeezing is achievable by increasing the feedback. Note that stability requires that $\eta$ should be not larger than $4C$.

\section{THE DETECTION OF THE MECHANICAL SQUEEZING}
In this section, we analyze that the mechanical squeezing can be measured by
the output field. The fluctuation
$\delta c(\omega)$ of the cavity field can
be obtained from Eq. (\ref{8}).
Using the input-output relation $c_{out}=\sqrt{2\kappa}c-c_{in}$ \cite{Walls}, we can get the fluctuation $\delta c_{out}(\omega)$ of the output field. Then we define the quadrature fluctuation of the output field as
\begin{eqnarray}\label{31}
 \delta z_{out}(\omega)&=&\frac{1}{\sqrt{2}}[\delta c_{out}(\omega)e^{-i\phi}+\delta c_{out}(-\omega)^{\dag}e^{i\phi}],
\end{eqnarray}
with $\phi$ being the measurement phase angle determined by the local oscillator. When $\phi=0$, $\delta z_{out}(\omega)=\delta x_{out}(\omega)$, which is the amplitude fluctuation of the output field. When $\phi=\pi/2$, $\delta z_{out}(\omega)=\delta y_{out}(\omega)$, which is the phase fluctuation of the output field.
Through calculations, $\delta z_{out}(\omega)$ is found to be
\begin{eqnarray}\label{32}
\delta z_{out}(\omega)&=&A_{z}(\omega)x_{in}(\omega)+B_{z}(\omega)y_{in}(\omega)\nonumber\\
& &+E_{z}(\omega)Q_{in}(\omega)+F_{z}(\omega)P_{in}(\omega),
\end{eqnarray}
where
\begin{eqnarray}\label{33}
A_{z}(\omega)&=&I(\omega)\cos\phi+R(\omega)\sin\phi,\nonumber\\
B_{z}(\omega)&=&R(\omega)\cos\phi+J(\omega)\sin\phi,\nonumber\\
E_{z}(\omega)&=&-\sqrt{\gamma_{m}}[A_{1}(\omega)\cos\phi+B_{1}(\omega)\sin\phi],\nonumber\\
F_{z}(\omega)&=&-\sqrt{\gamma_{m}}[A_{2}(\omega)\cos\phi+B_{2}(\omega)\sin\phi],\nonumber\\
I(\omega)&=&\frac{2\kappa}{d(\omega)}v(\omega)\Big\{|g|^2+[u(\omega)+2G\cos\theta]v(\omega)\Big\}-1,\nonumber\\
R(\omega)&=&\frac{4\kappa}{d(\omega)}G\sin\theta v(\omega)^2,\nonumber\\
J(\omega)&=&\frac{2\kappa}{d(\omega)}v(\omega)\Big\{|g|^2+[u(\omega)-2G\cos\theta]v(\omega)\Big\}-1.\nonumber\\
\end{eqnarray}
The spectrum of the quadrature fluctuation $\delta z_{out}(\omega)$ of the output field is defined by
\begin{eqnarray}\label{34}
S_{zout}(\omega)&=&\frac{1}{4\pi}\int_{-\infty}^{+\infty} d\Omega\ e^{-i(\omega+\Omega)t}[\langle \delta z_{out}(\omega)\delta z_{out}(\Omega)\rangle\nonumber\\
& &+\langle \delta z_{out}(\Omega)\delta z_{out}(\omega)\rangle].
\end{eqnarray}
Using Eq. (\ref{16}),
we find the spectrum of the quadrature fluctuation $\delta z_{out}(\omega)$ of the output field
\begin{eqnarray}\label{35}
S_{zout}(\omega)&=&[A_{z}(\omega)A_{z}(-\omega)+B_{z}(\omega)B_{z}(-\omega)](n^{th}_{c}+\frac{1}{2})\nonumber\\
& &+[E_{z}(\omega)E_{z}(-\omega)+F_{z}(\omega)F_{z}(-\omega)](n^{th}_{m}+\frac{1}{2}).\nonumber\\
\end{eqnarray}
 The output field is in a squeezed state if $S_{zout}(\omega)$ is smaller than that of the vacuum state, i.e., $S_{zout}(\omega)<\frac{1}{2}$.
\begin{figure}[htp]
\begin{center}
\scalebox{0.8}{\includegraphics{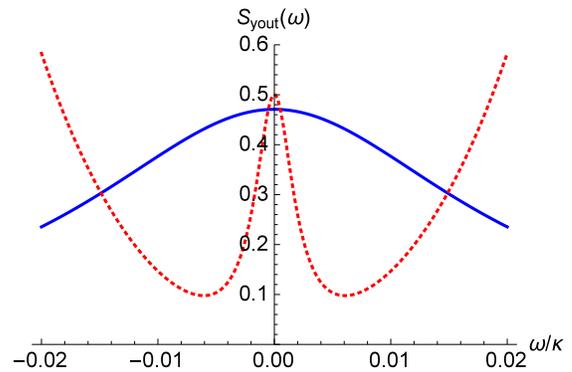}}\\
\caption{\label{Fig7}  The spectrum $S_{yout}(\omega)$ of the phase fluctuation of the output field
versus the frequency $\omega$ when $G=0.49\kappa$, $\theta=\pi/16$, and $T=0$ K in the absence of the optomechanical coupling $(g=0)$ (blue solid) and in the presence of the optomechanical coupling $(g\neq0)$ (red dotted). Here the spectrum $S_{yout}(\omega)$ for $g=0$ has been divided by 100.}
\end{center}
\end{figure}
We take $\omega_{m}/\kappa=10$, $\gamma_{m}/\kappa=10^{-5}$, $C=|g|^2/(\kappa\gamma_{m})=400$. The figure \ref{Fig7} plots the spectrum $S_{yout}(\omega)$ of the phase fluctuation of the output field
versus the frequency $\omega$ when $G=0.49\kappa$, $\theta=\pi/16$, and $T=0$ K without the optomechanical coupling $(g=0)$ and with the optomechanical coupling $(g\neq0)$. It is noted that the squeezing does not exist in the phase fluctuation of the output field for $g=0$ because of $S_{yout}(\omega)\gg0.5$. The squeezing exists in the phase fluctuation of the output field for $g\neq0$ since $S_{yout}(\omega)$ can be less than 0.5 when $|\omega|\leq0.0187\kappa$. Hence in the presence of the optomechanical coupling, the phase squeezing of the output field in $|\omega|\leq0.0187\kappa$ is a signature of the mechanical squeezing.
In the presence of the optomechanical coupling, the contour plot of the spectrum $S_{zout}(\omega)$ of the quadrature fluctuation of the output field versus the frequency $\omega$ and the phase $\phi$ when $G=0.49\kappa$, $\theta=\pi/16$, and $T=0$ K is shown in Fig. \ref{Fig8}. The lower figure in Fig. \ref{Fig8} indicates the region in which the quadrature fluctuation $\delta z_{out}(\omega)$ of the output field is squeezed. Therefore, the mechanical squeezing can be detected by measuring the quadrature fluctuation of the output field \cite{fbout}.
\begin{figure}[htp]
\begin{center}
\scalebox{0.7}{\includegraphics{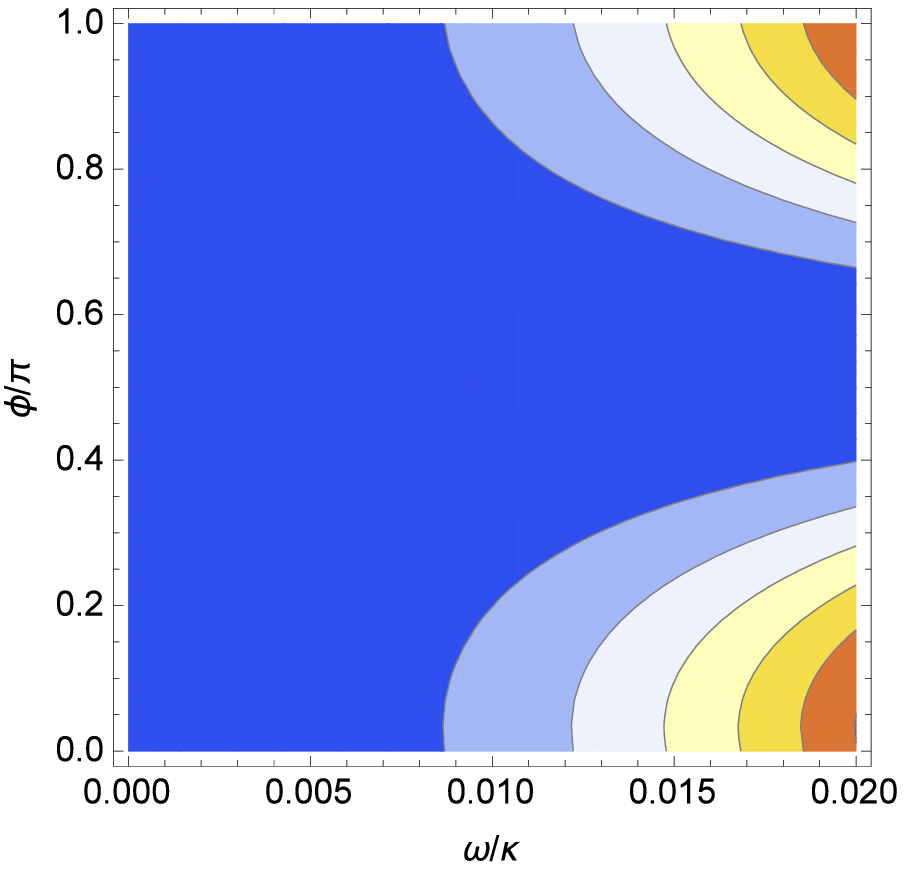}}
\scalebox{0.8}{\includegraphics{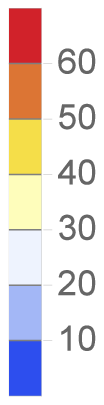}}\\
\scalebox{0.5}{\includegraphics{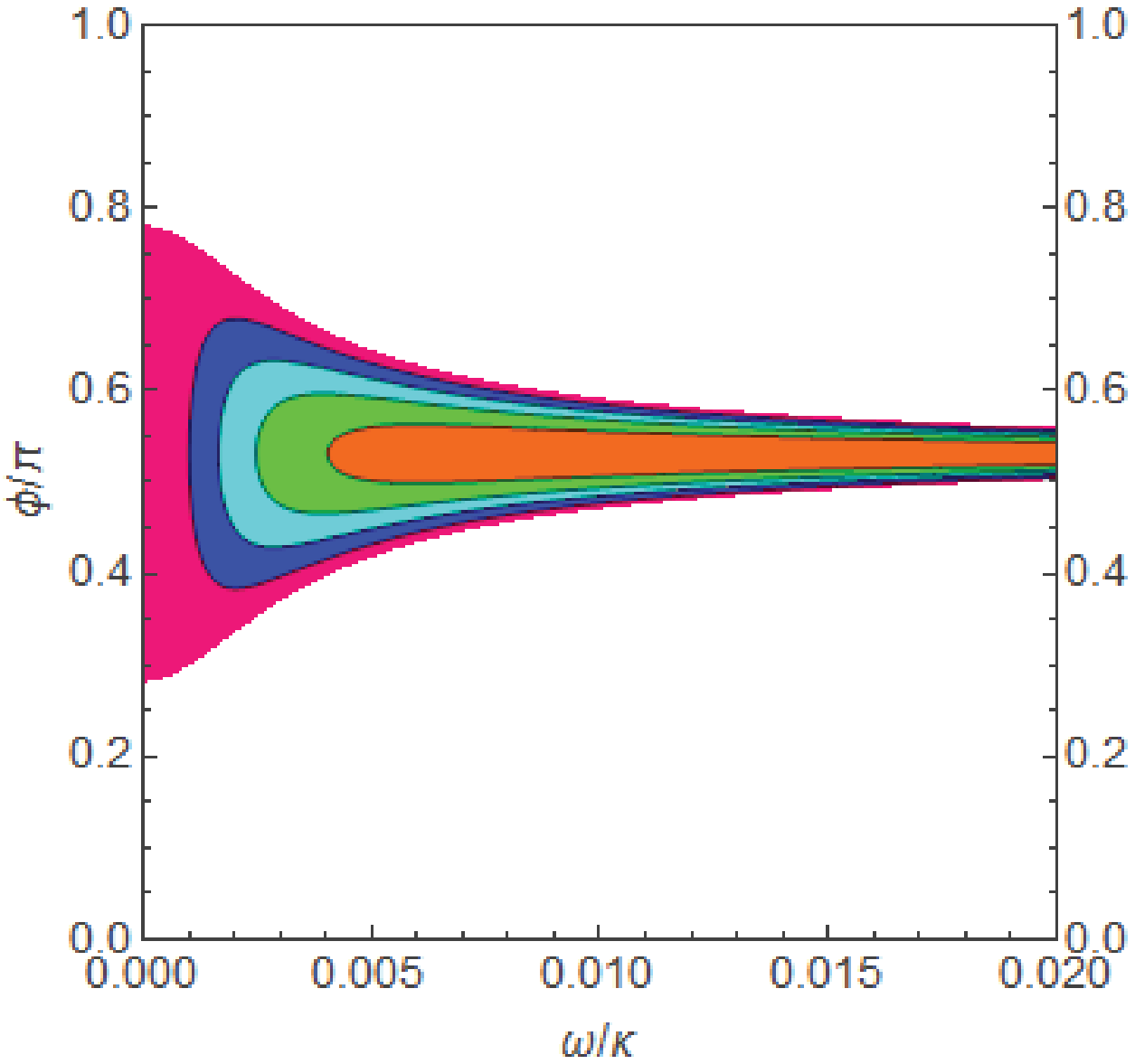}}
\scalebox{0.8}{\includegraphics{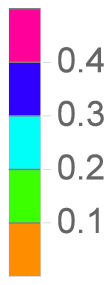}}
\caption{\label{Fig8}  The contour plot of the spectrum $S_{zout}(\omega)$ of the quadrature fluctuation of the output field
versus the frequency $\omega$ and the phase $\phi$ when $G=0.49\kappa$, $\theta=\pi/16$, and $T=0$ K. The lower figure zooms the upper figure for the region around zero.}
\end{center}
\end{figure}

\section{Conclusions}
In conclusion, we have demonstrated that the momentum fluctuation of the movable mirror can be squeezed when a PA is placed inside the cavity. It is found that the squeezing of the cavity field produced by the PA in the cavity can be fully transferred to the movable mirror in the resolved sideband limit and the thermal noise contribution is suppressed by a factor of the cooperativity parameter $C$. Moreover, we show that the detection of the mechanical squeezing can be realized by measuring the squeezing of the quadrature fluctuation of the output field by working in a regime of parameters when the PA does not squeeze the output field for no optomechanical coupling. In our work the degree of the
mechanical squeezing will be limited by the squeezing
produced by the PA. However one can use the previously
used methods like the single quadrature feedback scheme
\cite{Vinante} or the weak measurement \cite{Bowen,Pontin} to substantially
increase the mirror's squeezing as explicitly discussed at the end of Sec. V.

\setcounter{equation}{0}  
\section*{APPENDIX: THE SQUEEZING OF THE CAVITY FIELD IN THE ABSENCE OF THE OPTOMECHANICAL COUPLING}
\renewcommand{\theequation}{A\arabic{equation}}

For completeness and for making contact we present in this appendix what is well known for a cavity containing a PA \cite{Walls, Kimble}. In the absence of the optomechnical coupling $(g=0)$, the amplitude and phase fluctuations of the cavity field can be found from Eq. (\ref{8})
\begin{eqnarray}\label{A1}
\delta x(\omega)&=&A_{3}(\omega)x_{in}(\omega)+B_{3}(\omega)y_{in}(\omega),\nonumber\\
\delta y(\omega)&=&A_{4}(\omega)x_{in}(\omega)+B_{4}(\omega)y_{in}(\omega),
\end{eqnarray}
where
\begin{eqnarray}\label{A2}
A_{3}(\omega)&=&\frac{\sqrt{2\kappa}}{u(\omega)^2-4G^2}[u(\omega)+2G\cos\theta],\nonumber\\
B_{3}(\omega)&=&\frac{\sqrt{2\kappa}}{u(\omega)^2-4G^2}2G\sin\theta,\nonumber\\
A_{4}(\omega)&=&\frac{\sqrt{2\kappa}}{u(\omega)^2-4G^2}2G\sin\theta,\nonumber\\
B_{4}(\omega)&=&\frac{\sqrt{2\kappa}}{u(\omega)^2-4G^2}[u(\omega)-2G\cos\theta].
\end{eqnarray}
Without the PA in the cavity, $G=0$, we obtain $\delta x(\omega)=\frac{\sqrt{2\kappa}}{\kappa-i\omega}x_{in}$, $\delta y(\omega)=\frac{\sqrt{2\kappa}}{\kappa-i\omega}y_{in}$.
The spectra of fluctuations in the quadratures of the cavity field are defined by
\begin{eqnarray}\label{A3}
S_{Z}(\omega)&=&\frac{1}{4\pi}\int_{-\infty}^{+\infty} d\Omega\ e^{-i(\omega+\Omega)t}[\langle \delta Z(\omega)\delta Z(\Omega)\rangle\nonumber\\
& &+\langle\delta Z(\Omega)\delta Z(\omega)\rangle],\qquad Z=x, y.
\end{eqnarray}
With the help of the nonzero correlation functions of the noise sources in the frequency domain in Eq. (\ref{16}),
we obtain the spectra of fluctuations in the quadratures of the cavity field
\begin{eqnarray}\label{A4}
S_{x}(\omega)&=&[A_{3}(\omega)A_{3}(-\omega)+B_{3}(\omega)B_{3}(-\omega)](n^{th}_{c}+\frac{1}{2}),\nonumber\\
S_{y}(\omega)&=&[A_{4}(\omega)A_{4}(-\omega)+B_{4}(\omega)B_{4}(-\omega)](n^{th}_{c}+\frac{1}{2}).\nonumber\\
\end{eqnarray}
In the absence of the PA in the cavity, we have $S_{x}(\omega)=S_{y}(\omega)=\frac{2\kappa}{\kappa^2+\omega^2}(n_{c}^{th}+\frac{1}{2})$, which have peaks located at frequency zero with full width $2\kappa$ at half maximum.
The mean square fluctuations $\langle \delta x(t)^2\rangle$ and $\langle \delta y(t)^2\rangle$ in the quadratures of the cavity field are
determined by
\begin{eqnarray}\label{A5}
\langle \delta Z(t)^2\rangle&=&\frac{1}{2\pi}\int^{+\infty}_{-\infty} d\omega\ S_{Z}(\omega),\quad Z=x, y.
\end{eqnarray}
Without the PA in the cavity, we obtain $\langle \delta x(t)^2\rangle=\langle \delta y(t)^2\rangle=n_{c}^{th}+\frac{1}{2}$. For $T=0$ K, $n_{c}^{th}=0$, the cavity field is in a vacuum state, we have $\langle \delta x(t)^2\rangle=\langle \delta y(t)^2\rangle=\frac{1}{2}$.
According to the Heisenberg uncertainty principle,
\begin{eqnarray}\label{A6}
 \langle \delta x(t)^2\rangle\langle \delta y(t)^2\rangle\geq|\frac{1}{2}[x,y]|^2,
\end{eqnarray}
where $[x,y]=i$.
If either $\langle \delta x(t)^2\rangle$ or $ \langle \delta y(t)^2\rangle$ is below $\frac{1}{2}$, the cavity field is in a squeezed state. Similarly, the degree of the squeezing can be calculated by
$-10\log_{10}\frac{\langle \delta y(t)^2\rangle}{\langle \delta y(t)^2\rangle_{vac}}$ dB, where $\langle \delta y(t)^2\rangle_{vac}$ is the phase variance of the vacuum state and $\langle \delta y(t)^2\rangle_{vac}=\frac{1}{2}$.

\begin{figure}[htp]
\begin{center}
\scalebox{0.95}{\includegraphics{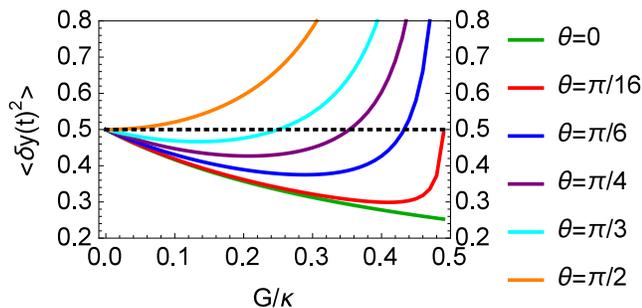}}
\caption{\label{Fig9}  The mean square fluctuation $\langle \delta y(t)^2\rangle$
versus the parametric gain $G$ for different parametric phases $\theta=0, \pi/16, \pi/6, \pi/4, \pi/3, \pi/2$ when $T=0$ K. The flat dotted line
represents the phase variance of the vacuum state $\langle \delta y(t)^2\rangle_{vac}=0.5$. }
\end{center}
\end{figure}

The numerical results show that $\langle \delta x(t)^2\rangle$ can not be less than $\frac{1}{2}$,
but $\langle \delta y(t)^2\rangle$ can be less than $\frac{1}{2}$. Thus we are interested in $\langle \delta y(t)^2\rangle$ here. The mean square fluctuation $\langle \delta y(t)^2\rangle$ as a function of the parametric gain $G$ for different parametric phases $\theta=0, \pi/16, \pi/6, \pi/4, \pi/3, \pi/2$ when $T=0$ K is shown in Fig. \ref{Fig9}. Note that $\langle \delta y(t)^2\rangle=0.5$ in the absence of the PA, thus the phase fluctuation of the cavity field is not squeezed. In the presence of the PA, it is noted that  $\langle \delta y(t)^2\rangle<0.5$ can be obtained except $\theta=\pi/2$. Hence, the phase squeezing of the cavity field can be achieved when a PA is placed in the cavity. The best squeezing happens at $\theta=0$ and $G=0.49\kappa$, at which $\langle \delta y(t)^2\rangle$ is equal to 0.253, the corresponding amount of the phase squeezing is about 49.4\%, and the degree of the squeezing is $-10\log_{10}\frac{0.253}{0.5}\approx2.96$ dB.


\begin{thebibliography}{99}
\bibitem{fbout} The squeezing spectrum of the output changes to our benefit if feedback is added to our model. These calculations are too long to be included here.
\end{thebibliography}
\end{document}